# Microgravity and Dissipative Granular Gas in a vibrated container: a gas with an asymmetric speed distribution in the vibration direction, but with a null mean speed everywhere

P. Evesque
Lab MSSMat, UMR 8579 CNRS, Ecole Centrale Paris
92295 CHATENAY-MALABRY, France, e-mail: **pierre.evesque@ecp.fr**

**Abstract:**
*The main topic of this paper (part 4) is the interpretation of data from extended simulations published in previous Poudres & Grains (see P&G 17, #1 to #18) concerning the dynamics of N equal-size spheres in a* 3d *rectangular cell excited along* Oz *in 0 gravity.*(N=100, 500, 1000, 1200, 2000, 3000, 4000, 4500). *Different Oz excitation kinds have been used (symmetric and non symmetric bi-parabolic, symmetric and non symmetric saw teeth, thermal wall). No rotation is included, dissipation is introduced via a restitution coefficient* e= −V'$_n$/V$_n$ , *where* V'$_n$ *and* V$_n$ *are the relative ball speed along normal to ball centres after and before collision. It is proved that the local speed distribution along z is fundamentally dissymmetric in most part of the cell while the mean local speed is* 0. *This demonstrates the inability of a model based on a thermal bath (with a single local temperature) to describe this dissipative-granular-gas-system, even when assuming that this temperature varies in space. The other (1-3) parts sum up few results obtained in the very low density regime.*

**Pacs # : 5.40 ; 45.70 ; 62.20 ; 83.70.Fn**

What a curious idea *a priori* to want to study granular materials in microgravity. But to look well, we notice at first that the manipulation of granular materials by the man is a necessity on earth/ground and that it will also become it in weightlessness condition during spatial, or even sidereal, journeys. How to make indeed to feed, without mixing powders, to grinding the flour, to breaking walnuts, then to sorting out residues … How make sure that the granular garbage left in wandering in a spaceship cannot block some command, and put in danger the survival of the expeditionary force? How to enhance reliability of the starting up of reactors with powders and impose a just proportion of combustive … How will it be possible to make the human being and plants with their pollens live together; how to allow the man to live with the other animals, and their waste and the excrements. All this will arise maybe one day in a program of intersidereal conquest; but before it, the man already plans to accost asteroids, to study their structure and their formation, then to extract certain resources from it (as metals, water,… ) useful for his survival in the space, or to bring back them on earth/ground.

Is it thus too early to speak about applied research in this domain? The implementation of a realistic system of exploitation of these new "ores" risks to require more of 20-25ans (the duration of patents) … The NASA is programming the development for a long time, by targeting if possible niches for fast profitability: the mechanical behaviour of the circles " granular media " weakly confined was a part of its first priorities during the flights of Columbia, because it conditions the future accostings.

For our part, we were interested in the management of grains and beads by vibrations. From the point of view of the applications, we know that such systems are very useful on earth/ground for the transport, the management and the handling of solids (vibrations are used to transport by vibrating strips, for compaction and the mixing of powders, for granular sorting, to put in fluidisation of blocked hopper…). Can we extrapolate these phenomena in microgravity and use them simply? To want to build an experiment on this subject is a wager in itself; it is necessary to conceive simple and reliable automatic protocols of filling and draining, without possibility of blocking; it is necessary to be able to modify the geometry of containers without that there is leak and flight of grains/beads, and all this with the minimum of human intervention … This project also has an underlying fundamental problem: if we shake periodically a closed container containing a small quantity of grains/beads, grains/beads are going to stir and to leave everywhere. Can we assimilate this system in a kind of gas of balls? Can we then use the classic statistical concepts of physics, stemming from the thermodynamics, to determine its characteristics? Is it necessary to overpass these concepts and to define other physical quantities, with new meanings? We were lucky to propose this study at the European Space agency ( ESA) and at the Centre National d'Etudes Spatiales (CNES) which funded it. Since then, we have been able to achieve certain number of experiments in sounding rockets, in a Chinese satellite and in the Airbus A300-0g of Novespace-CNES, and the other projects are in progress in a second Chinese satellite and in the orbiting International Space Station (ISS). I am going to describe some of the aspects.

I shall describe briefly the former(ancient) aspects (left 1-3) already developed in several other *Poudres & Grains* [1-5 ], to detail several recent results (part 4) obtained by numeric simulations which show that the real distribution of speeds in a gas granular media is never symmetric along the axis of vibration, although the local average speed is 0. *These two conditions cannot be together taken into account by the classic Boltzmann distribution. Here, we thus bring to light a specific behaviour of the dissipative granular systems which we must be able to find in the other experimental cases.*

## 1. A ball in Billiards:

First of all we noticed that a granular medium, contained in a box vibrated in weightlessness, has a "gaseous" behaviour (cf. Figure 1) that when the number density of its particles is neither too low nor too large, that means in fact the mean free average path $l_c$ of every grain/bead has to be of the order of the size L of the container.

We showed in a rectangular cell [1] that if the mean free path $l_c$ is much longer ($l_c > 10L$), balls do not almost meet; and trajectories remain linear, going throughout the cell; they return then the other way around by being reflected on the wall; balls go through practically the same road on the way out and on the way back if the geometry of the cell is simple (square, rectangle), because the lateral motion is blocked by the waste during the shocks with walls, and the movement becomes ***almost 1-dimensional***. Two cases appear then, according to the conditions of excitation: when the sinusoidal excitation is weak, the speed of balls is erratic, varying according to the conditions of the successive bounces; but it becomes almost constant in stronger amplitude, when the b/L ratio of the vibration amplitude b to the size L of the cell exceeds a certain value which depends on the restitution coefficient. In that case the movement of the ball thus simplifies itself enormously and becomes quasi-periodic and 1-d: we have then a totally regular system where balls(s) move in sync with the cell, going back and forth on a distance L in one or two periods [1]. Finally the third case is possible in theory, which shows all the complexity of this system, (but we have not verified it yet), it obtains when the restitution coefficient tends to the unity (elastic bounce); in that case, we show that the periodic motion of the ball loses its stability; the ball speed becomes again erratic from a collision to the other one; this case is not very probable with balls, because the energy loss is too big; hence we shall not that it is **this** solution which we have to obtain a priori for a "quantum" particle confined in a vibrated box, when energy is preserved during collisions.

To end it with the "1 ball" case, it is also necessary to consider more complex geometries of cell, with concave regions and the others convex, the "1 ball" case should be able to generate complex more erratic trajectories, see chaotic as in the case of the billiards of Sinai; this study is very promising because it touches the foundation of the theory of the chaos, but it still remains to make (as many of the other works in this domain).

## 2. Condensation and granular Maxwell daemon :

When the number of balls is too big, the dissipation takes it and the system is very difficult to excite experimentally. More, the motion is lived by a parasite noise we call the "g-jitter" or "noise of fluctuations in gravity"; this noise is the result of several factors, such i) the controlled or uncontrolled rotation of the device on itself (Airbus, rocket, satellite), such ii) a bad piloting of the machine, which undergoes a slow and random drift of the trajectory around the chosen elliptic trajectory, such iii) still the "incoherent" movement of an astronaut if the spaceship is lived. These noises are relatively important in the Airbus ($10^{-2}$ g/g=0.01-0.02); it can be very low in a rocket or a satellite ($10^{-5}$g); and it is intermediate in an inhabited orbiting space station ($10^{-4}$ g).

Typically, this noise takes it on the phenomenon to be studied as soon as the mean free path $l_c$ belongs of the order of L/10 to L/20 for a rocket. Naturally, the real value also depends on the level of excitation and on the loss by ball-ball collision, which we supposed here equal to 10-20 %. In an ideal experiment, as that obtained by simulation, balls should be probably never at rest and the simulations predict the same various phenomena such the existence of a kind of local condensation with an extremely localized dense system surrounded by a gas, but this appears that for $l_c$< L/20 and was not observed yet.

This condensation is interesting in the principle, because it will allow the manipulation of grains/beads and their transfer from a container to the other one: for example, let us consider a double container separated by a wall with a removable aperture; and let us consider first the complete cell without the separation wall; it contains a granular gas in dynamical equilibrium with a non uniform distribution (as we are afterward going to see it); however now we place the separation with the aperture in an adequate place, this one can force the "condensation" of the granular medium in the centre of one of the half-containers (say the first one), then this condensation will attract the quasi-totality of grains/beads going out of the second container; we shall so have made a differential pump with "sand" which will allow to activate a global migration of the system; this phenomenon works on earth/ground where it is baptized "Maxwell's Devil" of "granulars"  [2]. Will it work in microgravity? We hope for it, and the experiments are in the course of programming; they require long durations of experiments and a good quality of microgravity, which is the use of a satellite or an orbiting space station. But nothing is certain in this domain because it is little cultivated/studied. Everything also depends on real properties of granular gases (that we are now going to describe); now this study brought us its lot of surprises; this will allow us to exemplify why the numerical simulations are very often inefficient (even counter producing) if we do not associate them with a strict protocol so coherent and successful as those which the experimenters set up for their experiments: too often we content with identifying a relevant parameter and with studying its evolution according to the parameters of simulations; but are we always sure that this parameter is the voucher [2]? Should not we analyze at first the results in detail? This is because in any non linear complex system we risk certain number of surprise. It is what teaches us the physics of the disorder and that of soft material; it is what the management of companies learns also, which says to us that we can make economies of scale, or the process engineering, which teaches that the functioning of a pilot is not identical according to its size … But it is another problem.

### 3. Gaz granulaire, un cas « simple » de nano fluidique :

We are thus going to describe now some properties of the "true" granular gases, first of all from the typical behaviour propped up on our experimental results; then we shall remind the classic conventional description, which diverges appreciably from the previous presentation but which seemed to be supported by numerous numerical works. This decided on us to lead ourselves a series of simulations exposed there in § 4). Their analysis will show the inadequacy of the model and will confirm our understanding of the phenomena.

## *3.a. Résultats expérimentaux*

As we have already mentioned it, a simple granular gas can exist only in large dilution, when the collisions are "reasonably" dissipative (10-20 % losses per collision). This amounts in terms of mean free path $l_c$ between 2 collisions; for a granular gas containing N particles, we find the condition $l_c > L/10$ to $L/20$. This condition can be expressed i) in terms of real density ρ ($ρ=N/L^3$) and of size d of particles, thanks to the classic relation $ρl_c d² = 1$, or ii) in terms of number n of layers of grains/beads which recover the bottom of the container when balls are for rest and under gravity; indeed we obtain $l_c=L/(π n)$, because n=Nd²/L²; and we find then n < 3-6!

In these conditions, the gas is extremely rarefied and its mean mass density $4π ρd^3/3$ is small all the more as d is small; this implies that the physics of the system is not "extensive", that is it varies with the size of the system, with constant density: in other words, if we gather two identical systems in the same cell of double volume, the averaged behaviour in this bigger cell will be different from the behaviour in each of 2 systems taken remotely; this is thus very different from what is supposed to be for a gas or for a classic liquid for which one can define intensive quantities (pressure, temperature) which do not vary according to the size of the system and the others extensive quantities (volume, entropy, energy) which vary proportionally in the volume of the sample.

The regime for which the granular gas possesses a uniform pressure is the one for which grains/beads do not almost meet one another (that is $l_c > L$ or n < 0.3); it corresponds to the regime said of Knudsen for a real gas; this one obtains in cases of extreme rarefaction or nano-fluidics. It is however this type of regime that we can approach in a granular gas. We studied this case. The distribution of speed is there very remarkable [3] with an exponential tail [$p(v) =\exp(-v/v_o)$], far from the classic Boltzmann distribution. We shall not approach further this precise case here. We suggested interpreting this result by means of two models where the boundary conditions play an important role: one of the models proposes that the gas obeys the optimal disorder principle while being forced by boundary conditions to follow the "velostat" boundary condition (which is not thermostat). The second model considers that the gas is diphasic with molecules almost at rest and the others which gain some energy at each collision with walls until they return to collide with a ball almost at rest, which stops it [3 ].

If we increase the density, the system becomes inhomogeneous as Figure 1 shows it. The densification in the centre of the cell is small if $n_{mean} < 1$, but it grows very fast with $n_{mean}$; of more importance, this dense zone is much less mobile; one can thus call it "heap", although grains/beads are always free; this heap is taken in sandwich between two layers of "Knudsen" gas, very loose; mechanisms of "evaporation" and "condensation" of grains/beads by the heap insure the balance of the various phases.

Furthermore, we observe that the container gets looser from gas periodically during the motion (this is more visible on the cells of left and on the centre of Fig. 1 than in that of the right cell where depletion exists all the same, but much smaller). This indicates that the average speed of the grains/beads of the gaseous phase is lower than the maximum speed of the container; in other words, that the conditions of excitation are *supersonic*.

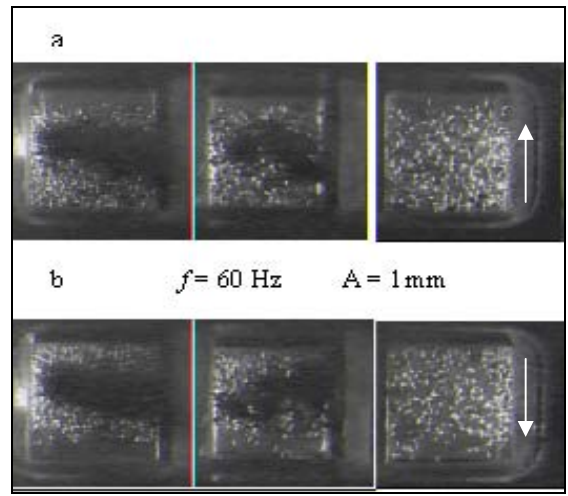

*Figure 1: balls vibrated in weightlessness; every cell is cubic ($10*10*10mm^3$) and contains a different number of grains/beads, from left to right, cell containing 3 layers of balls, 2 layers and 1 layers. Cells are at perigee (Fig a), and apogee (Fig b) and their motion is sinus in vertical in direction, with senses labelled by the arrow. Characteristics of the vibrations: frequency 60Hz, amplitude A=1mm, here.*
*We observe i) the formation of a dense heap surrounded with a gas in the case of the densest both cells, ii) the existence of a loose gas in the least dense cell, iii) the absence of the gas near the top wall of the top figure or of the bottom wall from the bottom figure, what shows that the excitation is "supersonic".*

All this allows noting the following:

*i) Mechanics of the discrete media –vs.-mechanics of the continuous media:* it is interesting to remind that the mechanical description of a gas (of atoms) from the equations of the mechanics of the continuous circles applies only for volumes bigger than the mean free path $l_c$. (For example, the equations of sound do not apply in the case of a gas of Knudsen, for which length $L<l_c$). The fact that a granular gas exists practically only in the regime of Knudsen lets augur some difficulties to describe it from the equations of the mechanics of continuous media.

*ii)* Existence of discontinuities: let us return to the "supersonic" character of the coupling between the gas and the wall; it is bound to the dissipative character of the ball-ball collisions which slows down strongly the internal dynamics of the gas. In terms of differential equations, this has to give rise to equations of the hyperbolic type, which admit discontinuities (shock waves), and not of the parabolic type (as in the case of the elasticity); in these conditions, it seems difficult to suppose the continuity of the variables everywhere in the medium. On the other hand, the Figure 1 does not show the existence of such shock waves propagating, and the heap seems continuous. How can we reconcile these view points?

*iii)     Problems of condition in borders and remark on the notion of temperature:*
Using the notion of granular temperature $T_g$ is perhaps dangerous. Indeed, this experiment shows that the important parameter which controls the speed V of particles is the speed $A\omega$ of the box; on the other hand, balls receive impulses from the walls. Must the box therefore be considered as a "velostat" and not as a thermostat? Is it a notable difference? Moreover the notion of thermostat calls the notion of energy and of thermal equilibrium; in our case contrariwise, walls inject a speed; therefore even if one consider that it is the energy of injection that is controlled, it is through the square of the speed; really injected energy therefore differs for gases with different particle masses $m_1$ and $m_2$ because they get different energy $½m_iV^2$. So, 2 systems of particles having the same number of layers will have the same speed V but not the same «granular temperature» $T_g=mV^2/2$; here come some difficulties to impose a local thermal equilibrium near the walls in case of tentative mixing, what can explain the phenomenon of segregation.

iv) *Difference between micro-gravity experiment and experiment on ground:* the fact that the granular gas cannot be completely described by equations of the mechanics of continuous media is difficult to bring to light on ground, because grains/beads are naturally confined by the gravity on the bottom of the container, so the interfaces vary abruptly on earth/ground. On the contrary, in microgravity, grains/beads collect in the centre; the density of grains/beads evolves and is small near walls what allows to study correctly the interface. Also on ground the heap which bounces rises up with a certain velocity V to a given height h, such as $½ V^2 \approx gh$, then falls again pushed by the gravity; the characteristic time $\tau=V/g$ is thus imposed by the gravity, this time is different to the one in 0g (where $\tau =L/V$).

v) Finally, we saw that the granular gas in weightlessness is obtained only for a number of layers n lower than 3-6; we can want to compare this problem with other phenomena arising on ground when the number of layers is very low (lower than 1.5); we fall then on granular Maxwell's daemon (that we quoted) and on that of the formation of oscillons, which are dynamic dissipative structures.

## 3.b.   *The simple model proposed by the literature:*

We are going to confront these results with our simulations, having called back the theoretical basis proposed usually in the literature. The basic model is to consider that walls act as a thermostat with temperature T [6,7]; these walls excite the granular medium locally; this excitement propagates step by step through the medium, by collision, while dissipating, because the collisions are dissipative. The local dissipation is reckoned from the local average speed <V> of the particles and the density $\rho$ of particles. We so obtain a distribution of density of particles $\rho(z)$ and of temperature T(z) similar to that of the Figure 2b. These two parameters vary the other way around one of the other one: the temperature is maximum near walls and minimum in the centre and the opposite for the density; more, the distributions are uniform perpendicularly in the direction of vibration (Oz) by reason of symmetry, if the ball-wall restitution coefficient equals 1.

In this model the distribution of speed is thus supposed to be very close to the local thermal distribution and has to obey the Boltzmann statistics, i.e. $p(v, z) = \exp-[mv^2/(2k_BT(z))]$, where $k_B$ is the Boltzmann constant, also represented in the Figure 2c; in this model the distribution of speed is thus almost isotropic locally.

The simulations [6], published in the literature, seem to confirm this model [7]. This one seemed to us however very different from what we found experimentally. This urged us to develop our own simulations.

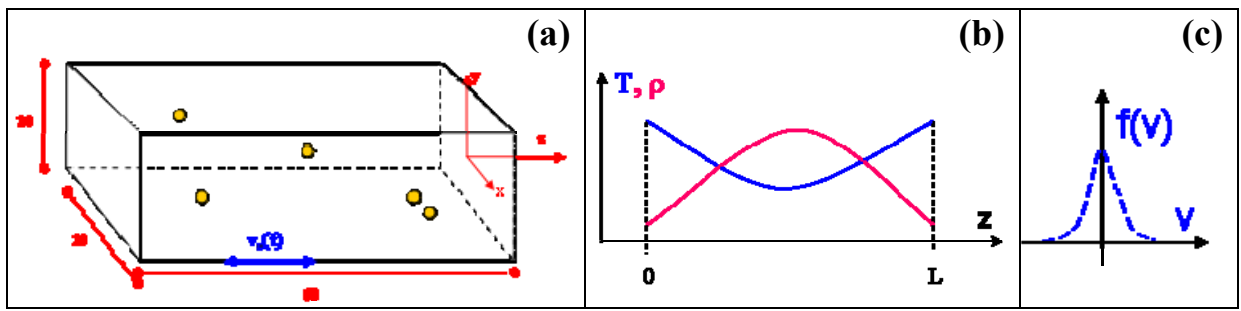

*Figure 2 : (2a) Scheme of the container, (2b) Distribution of the granular temperature T and of local density ρ predicted by the "thermostat" model along the axis of vibration (Oz), (2c) local distribution f(V) of the speed in all the directions; the width of the distribution is given by the local value of T.*

## 4. Our simulations :

As we said it, if most of the simulations seemed to confirm this model, this one appeared to us enough far from our results; we thus developed our own code of calculation and studied the behaviour of the system for various types of excitement (quasi-sine, triangular and thermal, cf. Fig. 3a)), for various numbers N of balls and for various coefficient e of restitution (e = - $V_{after}/V_{before}$, where $V_{after}$ (and $V_{before}$) stands for the relative ball-ball speed of approach before (or after) the shock. At first we did not want to introduce of rotation nor tangent forces and losses to simplify the study which contains already enough case (144 cases corresponding to 8 different numbers of balls (N = 100, 500, 1200, on 1600, 2000, 3000, 4000, 4500) * 6 excitation types * 3 coefficient e (e=0.7, 08, 0.9)). (We wanted to test in particular the effect of the various excitation forms because it seemed to us bizarre that they could not be correlated to the various distributions, as asserted it certain number of articles). With these simulations, we gave the means to determine the spatial distribution of balls, their distribution of speed in the complete cell and in all the directions… We can follow also the evolution of these distributions during the simulation to make sure that the system converges on a still dynamic state.

The totality of these results is published in previous articles this one [8], without comment. They were recently explained in several congresses the transparent of which we published [8b. We here want to explain clearly what does not agree in the previous model [7] (§-3.b). We shall thus be interested here only in the relatively dense systems (N > 800) for which a heap begins to form in the centre of the cell and for which the effects of excitement "supersonic" must be visible.

## *4.a. Distribution of $V_x$ or $V_y$ :*

In the particular case which interests us, the gas granular medium is rather dissipative. We shall thus suppose that the distribution of speeds depends on the place in the cell and we shall report it by drawing the distribution of speed of balls completely of this cell. However, we do not try to study the formation of heap in the direction perpendicular to z, what is verified as long as the density of grains(beads) is low(weak) (N < 5000 in our case).

We do find that in all our simulations [8], the distributions are homogeneous in layers at constant z. Furthermore, we observe that for the motions parallel to x or to y, the distributions p($V_x$) and p($V_y$) are symmetric with regard to the 0 speed value as indicated by the Figure 3b, and as the symmetry of the system imposes it. The width $\delta V_{x,y}$ of these distributions are linked to the granular temperature value in x and y directions $\delta V_x = (2k_B T_{x,y}(z)/m)^{1/2}$. Furthermore, we find $T_x(z) = T_y(z)$ as imposes it a symmetry. In that case, model and simulations are thus in accordance.

We shall not grow more before the study of the differences of speeds $V_x$ and $V_y$ Let us also give up studying the nature of the tail of the distributions of speeds. Are they really gaussiennes?

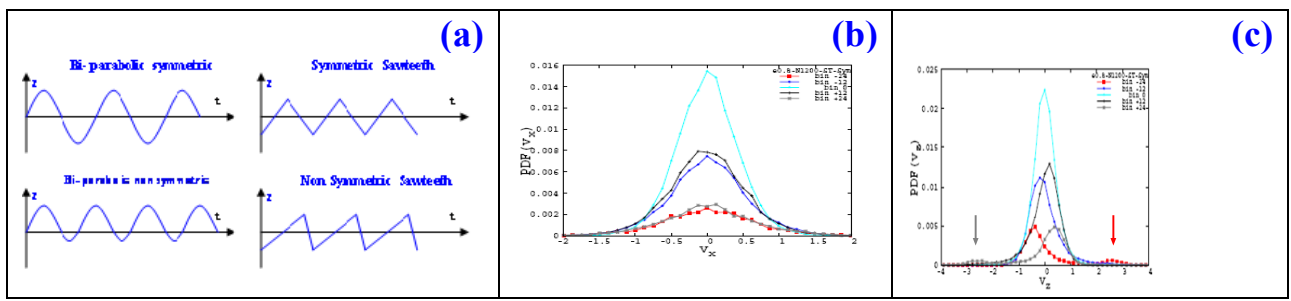

**Figure 3 : : Results of our numerical simulations for N=1 200 and a symmetric excitation in a box of 20d\*20d\*60d and a normal restitution coefficient e = 0.8: (3a) Various tested types of excitation. Results of the simulations obtained with a symmetric excitement (3b and 3c). (3b)** Distribution of the speed components $V_x$ or $V_y$ along Ox and Oy (i.e. perpendicular to the vibration Oz) for various positions z; these distributions are symmetric (red and grey curves: z = L/10 and 9L/10, near the side walls; curves blue and black: z=3L/10 and 7L/10 (middle-left and middle-right), blue-green: in the center z=L/2); these curves are symmetric. (3c) Distribution of the speed components $V_z$ along Oz (i.e. at the same time as the vibration Oz) for various positions z in the box; these curves are not symmetric (except at the centre of the box); We perceive a second peak for the very side positions; these peaks appear better in log-log coordinates) (cf. Fig. 4) (red: z=L/10, (near the left wall part of the cell); blue: z = 3L/10 (middle- left part of the cell); blue-green: z=L/2 (in the center of the cell); black: z = 7L/10 (middle-right part of the cell; grey: z =9L/10 (near the right wall).

### *4.b. $V_z$ Distribution :*

On the other hand we notice that the distributions of $V_z$ speeds in the Oz direction is not symmetric when ball approaches the edges z=0 and z=L of the container. This is explained by the transfer of impulse during the shock with walls: the particles which move slowly towards the mobile wall, go away from it then quickly after the shock. To display the effect of the wall excitation more exactly, it is better to use a saw tooth movement at speed V=constant as Figures 3c and 4a show it. In that case we see a single peak of speed centred around $V_z=0$ for balls in the centre of the cell (z=L/2); this peak, of amplitude $\rho_o$ for z=L / 2, decreases and excentres slightly on the right for balls in the right part of the cell (and mutually on the left for the balls of the left part); furthermore its width, $\delta V_z$, is very lower than $V_e$ and evolve according to z.

But, for the distributions close to one of the side walls (left z=0, or right z=L), we observe a new peak, weaker, which appears respectively above $+2V_e$ and below $-2V_e$. This peak, of width equal to the central peak $\delta V_z$, corresponds to particles returning in the centre of the cell, having bounced on the wall and having gained speed and energy; it is thus normal that the average speed of these new peaks is bigger than $|2V_e|$, and that the width of their distribution is equal to the width $\delta V_z$ of the thermal speed of the incoming slow particles; (here we always define the granular temperature as $\frac{1}{2}m(\delta V_z)^2/k_B=T_z(z=0\ ou\ L)$.

The position ($\pm 2V_e$) of this peak is such as $\delta V_z<<2V_e$ when the number of particles is big (more than a layer of balls), what demonstrates the "supersonic" character of the wall motion (as it is experimentally observed); finally this second peak of amplitude $\rho_{max}$ decreases in height as we consider closer and closer layers of the centre (z=L/2): it leaves $\rho_e$ in z=0 (or z=L)), then it is drowned in the tail of the slow-balls distribution. The reason is that these fast balls are slowed down by collision with slow balls of the "thermal bath" as they progress towards the centre of the cell; the amplitude of the second peak will be equal in $\rho_{max}=\rho_e/2$ when the distance of the layer is equal to the mean free path of balls outgoing from walls.

Then the general distribution will become again more symmetric past this distance and will not present more than a single bump centred on $V_z=0$, towards the centre of the cell (cf. Fig. 4a).

### *4.c. Braking of the symmetry of the speed distributions in the Oz direction:*

We also notice that 2 peaks (in the centre $\rho'_o$ and on the side $\rho_e=\rho_{max}$) near walls does not have the same height ($\rho'_o >> \rho_{max} =\rho_e$), so that the side peak can seem unimportant. In fact it is important otherwise, i.e. neglecting it would introduce an enormous dysfunction by breaking the matter preservation rule: indeed, the system is supposed to be in permanent regime (and de facto it is it), what imposes that the flow of particles crossing a plan at z given in a given direction (+z or -z) must be equal to the flow from the other way around, what imposes $\rho_o\delta V_z \approx \rho_e(2V_e+\delta V_z)$. The ratio of the peak ordinates (i.e. their amplitude) is thus the inverse ratio of their abscissas (i.e. the speed), because $\delta V_z$ is small. This indicates the existence of a symmetry breaking between both directions ±z, dissymmetry is imposed by the excitation mechanism. This symmetry breaking is not taken into account in any thermal model proposed in the literature.

This symmetry breaking is important, and it has few other consequences as we are going to see it, because it propagates up to the centre of the system: to demonstrate it, one can consider to decompose at any given location particles into two categories, the particles (+) going forwards, $V_z > 0$, and the particles (-) going inwards, $V_z < 0$. Let us consider then any fictitious plane parallel to the xOy plane and crossing the cell at some abscissa z; we can measure the local densities $N_+$ and $N_-$ of particles who are crossing this plane in the positive direction (+) or in the other one (-); still we can measure the average speeds $V_+$ and $V_-$ of these $N_+$ and $N_-$ particles, and the $P_+$ and $P_-$ pressures exercised by particles (+) and (-) when crossing this plane, and finally the temperatures $T_+$ and $T_-$ of both subsets.

It is what we made for all simulations, even if we do not represent in Fig. 4b and 4c that curves of temperatures and pressures from the case studied here. We so see that the thermodynamics balance $T_+ = T_-$, $P_+ = P_-$, $N_+ = N_-$, $V_+ = V_-$ is reached only in the centre of the cell. This demonstrates the disagreement between the real physics and the model of the literature described higher, because this model does not respect a basic rule of preservation imposed by the boundary conditions.

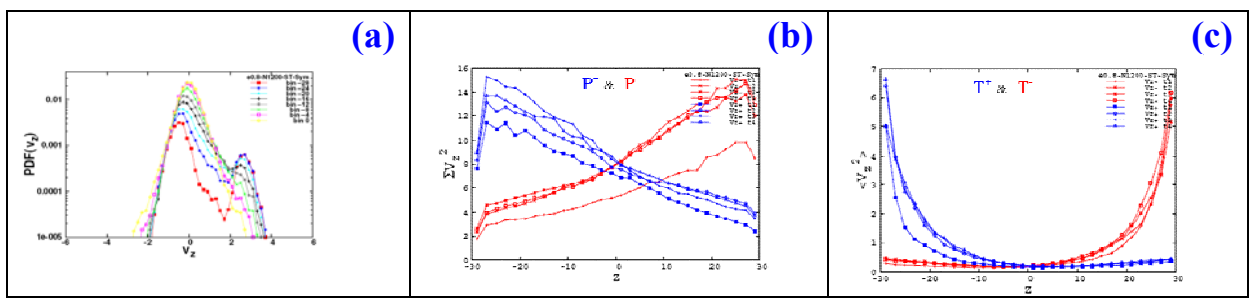

**Figure 4 : Results of our numerical simulations (same parameter as in the Figure 3, saw teeth excitation with N=1200, e=0.8):** (3a) The $V_z$ speed distributions at different z (cf. Fig. 3c), drawn in) log-linear coordinates: red curve: z=L/30, blue curve: z=L/10, blue-green: z=L/6, black z = 7L/30, grey: z=3L/10, green: z=11L/30, purple: z=13L/30, yellow: z=L/2. Results of simulations with symmetric saw teeth excitation (3b and 3c). (3b Distribution of the pressure components $P_+$ and $P_-$ by balls crossing a parallel plane in xOy from the left towards the right ($P_+$) and of the right towards the left ($P_-$), according to z. (3c) Distribution of the granular temperatures in the direction Oz for balls going towards +z ($T_+$) and -z ($T_-$). The 4 curves of $P_+$, $P_-$, $T_+$, $T_-$ which are displaid in (b) and (c) were obtained after 4 different times, what allows measuring the convergence of the dynamics towards a stationary state.

Naturally, we took here the most explicit case (sawteeth excitation), the one for which the wall speed is constant. What happens thus when we take different excitations (sine, double parabola with opposite apex, symmetric or asymmetric thermal conditions).

We remind that if N(V) dΩ Is the local density of particles with speed V in the volume N(V) dΩ, the previous parameters are given by partial integration over the positive (or negative) speeds, following the equations:

| | |
|---|---|
| $N_+ \, d\Omega = d\Omega \int_{V_z>0} N(V_z) \, dV_z$ | $N_- \, d\Omega = d\Omega \int_{V_z<0} N(V_z) \, dV_z$ |
| $F_+ = \int_{V_z>0} N(V_z) \, V_z \, dV_z$ | $F_- = \int_{V_z<0} N(V_z) \, V_z \, dV_z$ |
| $P_+ = \int_{V_z>0} N(V_z) \, (V_z)^2 \, dV_z$ | $P_- = \int_{V_z<0} N(V_z) \, (V_z)^2 \, dV_z$ |
| $T_+ = \int_{V_z>0} N(V_z) \, (V_z)^2 \, dV_z / N_+$ | $T_- = \int_{V_z>0} N(V_z) \, (V_z)^2 \, dV_z / N_-$ |

## *4.d. Other types of excitation:*

Fig. 5 give the results of the distributions $V_z$ obtained with the other types of excitation (thermal or sine). In this Fig. 5, as in the case of Fig. 4, we distinguish clearly an asymmetry of the distributions of speed $V_z$ in the left and right side parts of the cell, this as well in the case (a) corresponding to a sinusoidal excitation, as in the case (b) for a thermal excitation.
We notice finally an additional asymmetry in the case of Fig. 5b; this one is caused by a difference of "temperature of excitation" ($T_1 \neq T_o$) between left and right edges of the cell.

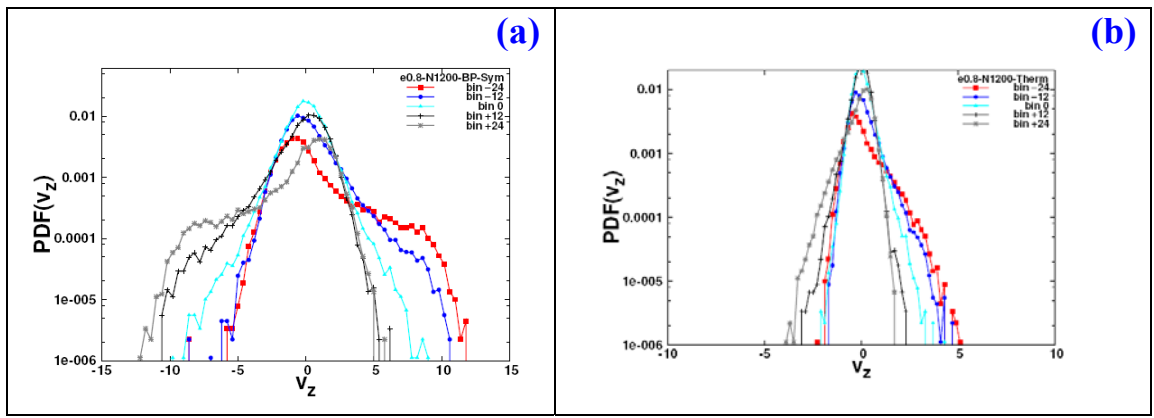

*Figure 5 : Distribution of speed $V_z$ according to the position z in the cell for (a) a bi-parabolic excitation (sine) and ( b ) thermal excitation. Total number of particles N = on 1200. Near edges (z=0, z=L) the distributions are not symmetric. The asymmetry is however weaker than in the figure 4. (b), both side temperatures of excitation are different ( $T_{right}=T_1=2T_0$), what gives a different asymmetry.*

Figs. 4 and 5 also show that the distributions of the local speeds depend on the type of excitation, i.e. sine, saw teeth, thermal. However, these variations are less perceptible on the averaged parameters such as average speed, local speed, or local temperature.

Fig. 6 give the distributions of the number $N_{tot}(z)$, $N_+(z)$, $N_-(z)$ of particles, of temperatures $T_+(z)$, $T_-(z)$ and of pressure $P_+(z)$, $P_-(z)$ for bi-parabolic symmetric excitations (a, b, c) on one hand, and thermal in 2 temperatures ($T_1=2T_0$) on the other hand. As in the case of Fig. 4, the asymmetry of the $V_z$ speed distributions is observed everywhere in the cell thanks to the distribution of the temperatures $T_+$ and $T_-$ and pressures $P_+$ and $P_-$, and this whatever is the kind of excitation. We can thus conclude that this speed asymmetry exists everywhere in the cell. This disagrees with the model proposed in the literature (§-3.b); besides this asymmetry is never taken into account in articles dealing with simulations of granular gases. In the best of the cases, articles give the speed distributions of averaged on the whole cell.

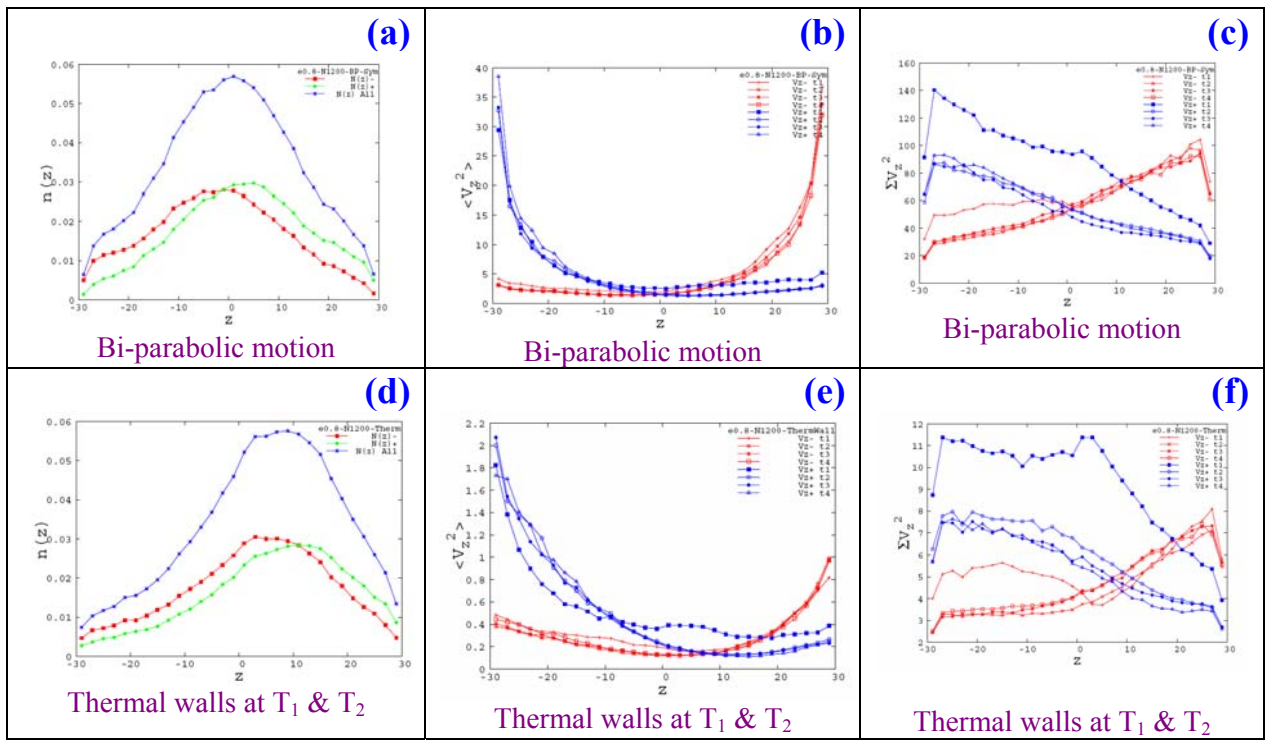

*Figure 6 : Distributions of the number $N(z)$, $N_+(z)$ and $N_-(z)$ of particles (a, d), of the temperature $T_+$ and $T_-$ (b, e) and of the pressure $P_+$ and $P_-$ for a symmetric bi-parabolic excitation (curves a, b, c) and for a thermal excitation with different temperatures $T_1$ and $T_2$ at the edge (curves d, e, f). The four curves of the same colour correspond to 4-different times of averaging and to different times of the simulation; they show the convergence of the simulation towards a still state.*

*Remark on the conditions of excitation at the walls:* we studied cases of symmetric excitation and others not symmetric, such as symmetric and not symmetric saw teeth, double top-to-tail symmetric and not symmetric parabolae. In the case of the thermal excitation we studied at the moment only the not symmetric case $T_1 = 2T_0$, cf. Fig. 6.

We did not try to resolve the case of a purely sine excitation, too expensive in calculation times. This was approximated by a wall motion having the shape of two portions of parabolas put top-to-tail; this allows to calculate times of collision exactly; the classic method of estimate [6] does not indeed allow to make exact calculations when the speed of particles is very lower than the typical wall speed [9].

### *4.e. Steady state :*

In Figs 4-6, we drew curves corresponding to four different times ($0-t_1$, $t_1-2t_1$, $2t_1-3t_1$ and $3t_1-4t_1$) after the beginning of the simulations. This allows studying/demonstrating the convergence of the calculation towards a steady state; the period of integration $t_1$ is constant for a total number of particles given, and curves thus give the evolution in time temporal averages (on $t_1$) of local number-density of balls, of local temperatures and local pressures. These curves are equivalent, if the state is steady.

We notice that the last curves on Figs. 4-6 tend well towards a common limit after a while, what demonstrates the final state is stationary. However in a certain number of cases, see [8], when N> 3000, curves drawn at different times continue to evolve, showing unsteady regime (cf. [8]) and proving that it is necessary to increase the integration time.

### *4.f. Discussion :*

Handling complex systems is often a necessity, because the world is complex. However, this example also shows how difficult it may be to get a clear understanding: who says complex system says large number of parameters, and thus large number of boring simulations; but it is not the more difficult it is then necessary to tidy up the results, that is often to introduce an additional "simple" idea which reports an incompatibility of the behaviour compared with average field.

To learn recognizing the inadequate cases is the base of the experimental methodology. It often requires an expensive systematic study which is the price to pay to avoid the errors, but which is the base of the development of our scientific and technical society; and to contribute to create, to promote and to maintain this tradition was the role of the big/main scientists.

In this precise case, the systematic study asks us to draw 6 000 different curves, then to introduce an unusual simplifying parameter and to demonstrate its efficiency. But without the respect for this procedure, the scientist questions the base itself of his know-how and of his own working reason.

In this case, many specialists of simulation found probably that it was to pay a high price too much, unless they missed idea to interpret the differences. In any case, to my knowledge, the published results were not interested in the local distributions, but indeed in the distributions, averaged globally on the whole sample.

We can see easily that if we do the same with the curves of Figs. 3 and 5a, the global distributions of speeds become again symmetric. Furthermore we notice that the results which we obtain from Fig. 3 compare correctly with those published in [10] (with the same series of parameters of simulation). Our results are thus compatible with those of the literature. But they describe them in a complete different manner.

Finally, it is necessary to notice that the same averaging procedure applied on of the distributions of Fig. 5b give curves which remain asymmetrical. The cause lies on the difference of excitation from both sides of the cell. This new asymmetry is not thus equivalent to the asymmetry that we put previously in evidence; this shows well the difficulty treating this kind of complex system.

**Future possible applications:** these results show that the real distribution of speeds in a dissipative granular gas may not be symmetric, although the local mean speed stays null. These two conditions cannot be together achieved using the classic Boltzmann distribution. We have thus just brought to light a specific behaviour of the dissipative granular gases. This one must be general and we must be able to find it in other experimental cases. The question is thus now where one has most chance to find it.

For example, it is most likely that it intervenes in many vibrated systems on earth/ground, although the revealing of the phenomenon is made more difficult in that case, because the granular medium is compressed by the gravity. This makes difficult a local study of the distributions. However, one can look for it in granular gases under vertical or horizontal vibration, in the "Leidenfrost granular" effect, or in the case of oscillon generation …

Besides, one should envisage the possibility of finding this type of phenomenon in the rapid granular flows. Indeed, what shows this simulation is the existence of dissipative structures which are not at thermodynamics equilibrium and which generate several species in collision interaction (here 2 species (±), but why not 3 or 4). This type of phenomenon could then be engendered by fast flows with large variation of density.

## 5. Conclusion :

The mechanics of the granular media in microgravity is thus often a source of new results, and the real behaviour is different from the expected one. For example we had not planned to linger over the case of a single ball; and the experiment was achieved to calibrate the restitution coefficient; this behaviour was not described in the literature, it amazed us; from there was born a more precise method to measure the restitution coefficient in the absence of rotation.

Also the behaviour of the granular gas which we described is very different from flows in the thermal equilibrium. The local asymmetry between balls with positive and negative speeds is a consequence of the excitation; it is not reducible in an average; it propagates in the whole sample: every category has its own temperature, different from the other one except at a point in the cell (here the centre); it provokes a local difference of pressure $P_+ \neq P_-$ which maintain the system in a denser and calmer state in the centre than near the walls. Everywhere, balls look to belong to one of the two different categories; nevertheless balls can pass from a category to the other one by simple collision; they are thus exchangeable, what means that we cannot report the system by introducing simply two classes of particles (we should give then also the transmutation conditions). Finally, the local speed, averaged over the local sum of two classes, is null (everywhere of the system), because the mechanical state is steady; The distribution of speed of the set formed by both classes thus has a null average (i.e. flow or average speed equals 0 everywhere), but the standard deviation (related to $T_+$) of the positive wing of the distribution is different from that ($T_-$) of the negative wing. The distribution is not thus Gaussian and it presents a fundamental asymmetry which increases with the distance fro the cell centre, as if the entities + and - resulted from two different systems. It is the reason for which the "thermal" model proposed by the literature, and described briefly in sub-section 3b, cannot be correct.

In conclusion, it seems to me that the characteristics of the phenomenon which we have just characterized is very far from the classic formulations for equilibrium, and requires probably to introduce mathematical tools and special concepts.

This is a summary of the first steps of this discipline. It shows the complexity of the behaviour of the granular media, as well as its originality. Many other results must be obtained to allow the man to use the granular material in microgravity as he uses on earth/ground; let us quote for example the problems of mixing and segregation, of fluidisation, of convection, of transport, or the problems of impacts or grinding, compaction, see of combustion…

*Acknowledgements:* CNES, CNSA, ECP, ESA and IOP-CAS are greatly thanked for partial funding. I asked R.Liu to perform the numerical simulations during his 1-year stay (2008-2009) at ECP ; this stay was supported by China grant. The results are presented in the article [8]. The simulations have been redone and results have been confirmed by M. Chenna & D. Roddrigues, two license-students at Paris 11 Univ., during their stay in the lab in January 2010, then by E. Urania, a second year student at ECP.  I thank S. Luding for giving to me the opportunity to present these results and to give my  interpretation at the Powders & Grains meeting, July 2009. I thank M. Hou for inviting me in May-June 2010 in Beijing to participate in the jury of Liu's Viva/Defense. I was surprised that our results [8] did not appear in Liu's PhD dissertation [11]; but he is now preparing a paper about this subject.

## Appendix : Simulation technique

A program of molecular dynamics working in C has been used to simulate the dynamics of a colliding gas of equal spheres with dissipation, with equal mass m. Ball-ball collision is treated using inelastic restitution coefficient $e=v_f/v_i$ (=0.9, 0.8 or 0.7), excluding rotation effects and rotation parameters. Ball diameter D is the space unit (D=1). Rectangular box is used with dimension (x,y,z) = (20*20*60). Oz is along vibration; Transverse directions are Ox and Oy, no transverse motion of the box is imposed.

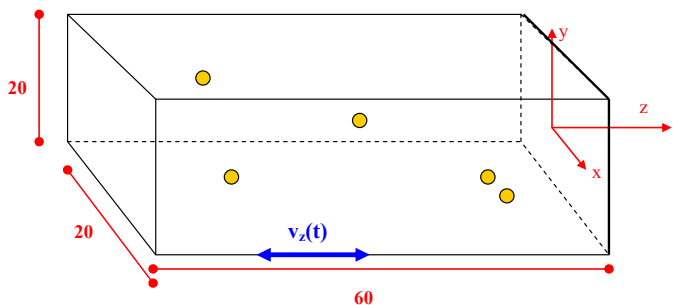

*(a)The shape of the container*

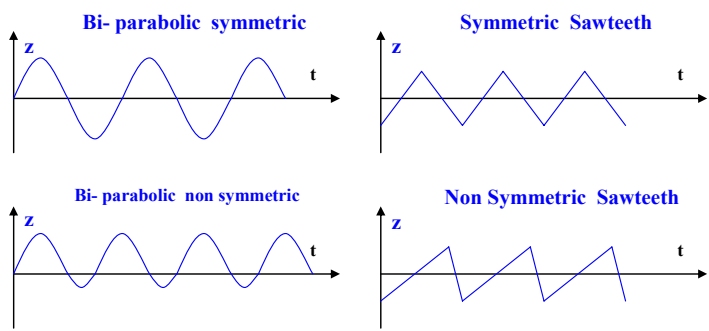

(b) Different excitation types of the vertical walls

**Figure symbols and abbreviations:**
e0.9: coefficient of restitution $e = 0.9$     N***: number of particles $N = $ ***     BP: bi-parabolic driving
ST: saw-tooth driving     Sym: symmetrical driving     Nsym: Non-symmetrical driving

We study 3d dynamics of N spheres (N=100, 500, 1200, 2000, 3000, 4000, 4500) with different excitation (symmetric and non symmetric bi-parabola and sawteeth drivings, thermal excitation ($\exp(-v^2/kt)$). In thermal excitation, balls which collide with moving wall get a random distribution according to the thermal noise. In bi-parabolic driving, the wall speed is assumed continuous and acceleration $+\Gamma_1$ is applied during $T_1$, then changes to $-\Gamma_2$ during $T_2$ and conversely; so a period $T=T_1+T_2$, and the continuity condition leads to $\Gamma_1 T_1 = \Gamma_2 T_2$. This excitation is quite similar to a symmetric sinus wave when $\Gamma_1 = \Gamma_1$.

The program finds ball-ball and ball-wall collisions and the snapshots of ball positions and speeds are recorded every (N/10) collisions; The program stops after 100*N collisions and contains 1000* snapshots of 3d- cell and balls. Steady state is obtained after some time. The cell is cut into 59 bins perpendicular to vibration direction, and the different local quantities are averaged over two consecutive bins.

Dynamics is studied in displaying different parameters such as the **p**robability **d**istribution **f**unctions (**pdf**) of the speed coordinates $V_z$, and $V_x$ (along and perpendicular to excitation respectively) at different position z, the density distribution n(z), the speed distribution $V_z(z)$ as a function of the position z, the mean speed $<V_z> = \Sigma_{particles} m V_z / (\Sigma_{particles} m)$, which is also the mean flow, the mean temperature $kT/m = \Sigma_{particles} V^2_z / (\Sigma_{particles})$ and the mean pressure $P_z = \Sigma_{particles} m V^2_z$. Only normal restitution coefficient e is introduced to take account of dissipation; No rotation and friction is included.

We also separate the particles into two sets at a given instant, *i.e.* those ones which move towards $z^+$ (positive $V_z$), and those ones which move towards $z^-$ (negative $V_z$) and we plot the same quantities with respect to these directions, i.e. the density distribution $n^{(\pm)}(z)$, the speed distribution $V_z(z)$ as a function of the position z, the mean speed $<V^{(\pm)}_z> = \Sigma_{particles} mV^{(\pm)}_z / (\Sigma_{particles} m)$, which is also the mean flow in + or - z, the mean temperature $kT/m = \Sigma_{particles}(V^{(\pm)}_z)^2 / (\Sigma_{particles})$ and the mean pressure $P_z = \Sigma_{particles} m(V^{(\pm)}_z)^2$, on graphs.

## References

[1] **billiard :** P.Evesque, *Poudres & Grains* **13**, 40-73 (2002), « Quelques Aspects de la Dynamique des Milieux Granulaires » ; http://www.mssmat.ecp.fr/IMG/pdf/poudres13_4-dyn.pdf ; P. Evesque , *Poudres & Grains* **12**, 17-73 (2001), « The Thermodynamics of a Single Bead in a Vibrating Container » ; P. Evesque , *Poudres & Grains* **14**, 8-53 (2004) (voir Appendice de "New corner stones in dissipative granular gases : On some theoretical implication of Liouville's Equation in the physics of loose granular dissipative gases"); P. Evesque, F. Palencia, C. Lecoutre-Chabot,  D. Beysens and Y. Garrabos, "Granular gas in weightlessness: the limit case of very low densities of non interacting spheres", ISPS 2004 (Toronto- 23-27 may 2004); *Microgravity Sci. Technol.* XVI-1, 280-284 (2005); M. Leconte, Y. Garrabos, F. Palencia, C. Lecoutre, P. Evesque, D. Beysens, "Inelastic ball-plane impact: An accurate way to measure the normal restitution coefficient", *Appl. Phys. Lett.* **89**, 243518  (2006).

[2] **Démon de Maxwell**: J. Eggers, "Sand as a Maxwell demon", *Phys. Rev. Lett.* **83**, 5322-25, (1999); J. Javier Brey, F. Moreno, R. Garcıa-Rojo and M. J. Ruiz-Montero, "Hydrodynamic Maxwell Demon in granular systems", *Phys. Rev.* **E 65**, p. 11305 (2001) ; P. Jean, H. Bellenger, P. Burban, L. Ponson & P. Evesque, "Phase transition or Maxwell's demon in Granular gas?", *Poudres & Grains* **13** (3), 27-38 (juillet-Août 2002); R. Mikkelsen, K. van der Weele, D. van der Meer, M. van Hecke and D. Lohse, "Small-number statistics near the clustering transition in a compartmentalized granular gas", *Phys. Rev.* **E 71**, p. 41302 (2005); A. Barrat & E. Trizac, "A molecular dynamics "Maxwell Demon" experiment for granular mixtures", ArXive:Cond-mat/0212054v1 (dec 2002) ; P. Evesque, " How one can make the bifurcation of Maxwell's demon in Granular Gas Hyper-Critical ", *Poudres & Grains* **16** (1), 1-20 (Février 2007); P. Evesque, "Cyclic Maxwell Demon in granular gas using 2 kinds of spheres with different masses" *Poudres & Grains* **16**, 23 (2007)

[3] **gaz granulaire avec n <1:** M. Leconte, Y. Garrabos, E. Falcon, C. Lecoutre-Chabot, F. Palencia, P. Evesque, D. Beysens, "Microgravity experiments on vibrated granular gas in dilute regime: non classic statistics", *Journal of Statistical Mechanics: Theory and experiment,* P07012 (2006); P. Evesque, Y. Garrabos, C. Lecoutre, F. Palencia, and D. Beysens, in *Powders & Grains 2005*, (Garcia-Rojo, Herrmann, McNamara ed., Balkema 2005), pp. 1107-1111 ; P. Evesque , *Poudres & Grains* **14**, 8-53 (2004)

[4] **gaz granulaire avec n <1 :** P. Evesque, *Poudres & Grains* **15**, 1-16 (2005); P. Evesque, *Poudres & Grains* **15**, 18-34 (2005); P. Evesque, "A model of dissipative granular gas: the ultimate case of complete inelasticity of grain-grain collision", *Powders & Grains 2005*, Stuttgart, July 18-22, 2005,in *Powders & Grains 2005*, (Garcia-Rojo, Herrmann, McNamara ed., Balkema 2005), pp. 1131-1134;

[5] **gaz granulaire, 1≤n<10 :** E. Falcon, R. Wunenburger, P. Evesque, S. Fauve, C. Chabot, Y. Garrabos & D. Beysens; *Phys. Rev. Lett.* **83** (12 juillet 1999) 440-443 ; P. Evesque: Comparison between Classical-Gas behaviours and Granular-Gas ones in micro-gravity : *Poudres & Grains* **12**, 60-82 (2001); P.Evesque, *Poudres & Grains* **13**, 40-73 (2002) ; P. Evesque , *Poudres & Grains* **14**, 8-53 (2004) ; P.Evesque, *Poudres & Grains* **16**, 38-62 (2007);  P. Evesque , A. Garcimartin, D. Maza Ozcodi, N. Vandewalle, Y. Garrabos, C. Lecoutre, D. Beysens, X. Jia, M. Hou ; (JASMA); J. Jpn SocMicrogravity Appl. **25**, 447-452 (ou 623-628) (2008); M. Hou & P. Evesque , In *Advances in Microgravity Science* , edited W.R. Hu research Signpost  (Research Signpost, Transworld Research Network, Keralda, India, 2008)

[6] T. Poschell & S. Luding, *Granular Gases,* Lectures Notes in Physics **564**, (Springer-Verlag, Berlin, 2001); *Granular Gas Dynamics*, Lectures Notes in Physics **624**, edited by T. Poschel and N. V. Brilliantov, (Springer-Verlag, Berlin, 2003); A. Barrat, E. Trizac & M.H. Ernst, "Granular gases: dynamics and collective effects", arXiv:cond-mat/0411435 v2, 3/12/2004, published in J. Phys. C (2005); S.Luding, R.Cafiero, H.J. Herrmann, "Driven Granular Gas", in *Granular Gas Dynamics*, Lectures Notes in Physics 624, edited by T. Poschel and N. V. Brilliantov, (Springer-Verlag, Berlin, 2003), 293

[7] J. Javier Brey, F. Moreno, R. Garcıa-Rojo and M. J. Ruiz-Montero, "Hydrodynamic Maxwell Demon in granular systems", *Phys. Rev.* **E 65**, p. 11305 (2001). I. Goldhirsch, "Rapid granular flow", *Annu. Rev. Fluid Mech.* **35**, 267 (2003) ;

[8] R. Liu, M. Hou, P. Evesque, *Poudres & Grains* **17** (1-18), 1-561 (2009) ;
   [8b] P. Evesque, R. Liu, M. Hou, *Poudres & Grains* **17** (19), 563-576 (2009) ;
   [8c] P. Evesque, *Poudres & Grains* **17** (20), 577-595 (2009) ;

[9] P. Evesque, "Boundary conditions and the dynamics of a dissipative granular gas : slightly dense case", *Poudres & Grains* **16** (3),38-62 (2007)

[10] W. A. M. Morgado & E. R. Mucciolo; Numerical simulation of vibrated granular gases under realistic boundary conditions; arXiv:Cond-Mat/0204084v1 (2002); W. A. M. Morgado and E. R. Mucciolo, *Physica A* **311**, 150 (2002)

[11] R. Liu, "Condensation and oscillation in dissipative granular gases", PhD thesis, CAS, IOP, Beijing, (May 28, 2010)

# Notice pour les auteurs de
# *poudres & grains :*

## Objet de la publication

***Poudres & Grains*** est une revue publiant des articles scientifiques originaux dont le sujet traite des matériaux en grains, en poudre ou assimilés; elle est couverte par le copyright. Elle s'adresse à des professionnels de la recherche et de l'enseignement des secteurs public et privé. Chaque numéro a une version imprimée conservée à la Bibliothèque de France. La reproduction intégrale des articles et/ou de la revue pour des usages personnels ou afin d'archivage est autorisé et peut se faire par téléchargement. Une autorisation doit être demandée pour des reproductions même partielles.

**Soumission des articles:** Les articles doivent être des originaux; un transfert de copyright doit être signé, spécifiant que l'auteur accepte les règles éditoriales, surtout celles relatives aux commentaires scientifiques, car les articles sont ouverts à discussion scientifique. Des liens électroniques seront établis dans la mesure du possible.

## Règles éditoriales:

**Tout auteur scientifique doit**
- Décrire honnêtement les résultats qu'il a obtenus tant théoriques qu'expérimentaux.
- Accepter et favoriser le débat honnête entre scientifiques.
- Ne pas faire de querelles de personne.
- Respecter les droits des autres auteurs scientifiques et de l'antériorité scientifique en particulier.

*Tout manquement à ces règles supprime l'accès à la publication. L'auteur est seul responsable du contenu de l'article. Une commission éditoriale donne son avis au besoin; mais le vrai travail de rapporteur doit être exécuté a posteriori après publication, par la communauté scientifique.*

**Tout lecteur scientifique doit**
- Faire une analyse critique des articles scientifiques qu'il lit de manière à se forger sa propre opinion
- A la suite d'une lecture scientifique, porter à la connaissance des lecteurs scientifiques des résultats faisant partie du domaine public et exprimant les mêmes résultats ou des résultats contraires à ceux qu'il vient de lire.

## Notice de Mise en Page et Règles Typographiques

La langue de la revue est le **français** ou **l'anglais**. **Longueur maximum de l'article:** 20pages; **format A5, lisible par Acrobat reader (format pdf).**

**marges:** haut: 1cm ;    bas: 1cm ;    gauche: 1.25cm ;    droite: 1.25cm;
reliure: 0cm;    entête haut:0.9cm;    entête bas: 0.9cm

**Polices:** police général: times new roman 10   ou   symbol 10
police pour les références, les légendes des figures, le résumé (abstract): times 8
abstract en italique sauf pour les caractères "symbole" ou en exposant ou en indice
police du nom des auteurs: times new roman 13
police du Titre: times new roman 13

**paragraphes:** en interligne simple

**ligne d'Espacement entre les paragraphes:** Général: 6pts
Que ce soit (i) entre 2 sections, (ii) pour la 1$^{ère}$ ligne du texte, (iii) avant, après ou entre équation.

**Retraits de la première ligne d'un paragraphe:**
Pas de retrait après une entête;
Autrement retrait de 0.75 cm

**Entête de page** contient 1 ligne avec à gauche l'Auteur et un titre abrégé (en italique, times new roman 8); à droite le n° de page en times new roman 10 , puis 1 ligne vide de caractère, taille 8pts.
**Exemple:** *A.Parson et al./ short title*                                                                                                       - 1 –

**Pied de page** contient 1 ligne vide de 6pts,
puis 1 ligne contenant le nom de la revue (*poudres & grains*) , son numéro (en gras times new roman 8), les numéros de pages et la date de parution. Attention le p et le g sont en times new roman 10 italique
**Exemple:** *poudres & grains* **6**, 10-16 ( août-septembre 1999*)*

**Adresse:** ne pas oublier de donner l'e-mail

**Couverture:** times new roman 12   (en gras italique pour le n° et la date; en italique pour le nom et la page).

**Adresse:**    P. Evesque, éditeur,        33 1 41 13 12 18; fax: 33 1 41 13 14 42; E-mail: evesque@mssmat.epc.fr
Lab MSSM, Ecole Centrale Paris, 92295 Châtenay-Malabry, France; **web**: prunier.mms.ecp.fr